\documentclass[wssquare]{ws-procs9x6}            
\usepackage{epsfig}           
\usepackage{epstopdf}

\usepackage{amssymb}
\usepackage{amsmath} 
\usepackage{amssymb} 

\usepackage{subfigure}

\usepackage{xspace}
\newcommand{\nubare}{\ensuremath{\overline{\nu}_e}\xspace}
\newcommand{\nubarenubare}{\ensuremath{\nubare \rightarrow \nubare}\xspace}

\usepackage[left]{lineno}

\begin{document}
\title{Solar neutrinos with the JUNO experiment}

\author{G. Salamanna$^*$ {\it on behalf of the JUNO Collaboration}}

\address{Roma Tre University and INFN Roma Tre, 00146 Rome, Italy\\
$^*$E-mail: salaman@fis.uniroma3.it}

\begin{abstract}
The JUNO liquid scintillator-based experiment, construction of which 
is on-going in Jiangmen (China), will start operations in 2020 
and will detect anti-neutrinos from nearby reactors; but also solar 
neutrinos via elastic scattering on electrons.
Its physics goals are broad; its primary aim to measure the 
neutrino mass ordering demands to collect large statistics, which 
requires JUNO's 20 kt sensitive mass, 
and achieve an unprecedented energy resolution (3$\%/\sqrt{E}$). 
Thanks to these characteristics, JUNO is in a very good position 
to contribute to the solar neutrino studies in the line of previous experiments 
of similar technology. 
It will collect a large sample of neutrinos from $^7$Be and 
$^8$B. In particular, for $^7$Be the target energy resolution will provide 
a powerful tool to isolate the electron energy end point from 
backgrounds like 
$^{210}$Bi and $^{85}$Kr. 
At the same time, challenges will have to be faced mainly related to the 
reduction and estimation of the backgrounds. 
While a thorough LS purification campaign is being planned, the 
desired level of purification is less aggressive than e.g. in Borexino. Also,
cosmogenic backgrounds such as cosmic ray muons 
traversing the relatively thin layer of ground above
JUNO (700 m) and 
crossing the detector will need to be vetoed with dedicated techniques for the 
extraction of $^8$B. 
In my talk I reviewed JUNO's preliminary analysis strategy and challenges in the 
solar neutrino sector; and provided the current estimates of its solar neutrino and background yields, with related energy spectra, assuming two 
benchmark scenarios of scintillator radio-purity. \end{abstract}


\bodymatter

\section{Introduction}
The Jiangmen Underground Neutrino Observatory (JUNO) is a neutrino 
experiment being built in China, described in Ref.\cite{An:2015jdp}. 
Its primary purpose is to determine the neutrino mass hierarchy (MH) 
and measure the oscillation parameters using reactor sources. It will
consist of a large mass of pure liquid scintillator (LS), placed in an acrylic 
sphere of 35.4 m of diameter; a system of 
large-area photo-multipliers of new generation (PMT); a veto system.
It will be located at a distance of approximately 50 km from two nuclear power plants
 (Yangjiang and Taishan). The two plants are expected to provide
an equal thermal power of about 18 GW, although at the start-up of 
the experiment only 26.6 GW are expected to be available. 
The baseline is optimized for maximum sensitivity to the mass hierarchy
determination driven by the survival oscillation probability
(here in the case of normal neutrino mass hierarchy (NH)): 
\begin{eqnarray}\label{eq:1}
P_{NH}&(\nubarenubare)=&1-\frac{1}{2}\sin^2 2\theta_{13}\left(1-\cos\frac{L\Delta m^2_{atm}}{2E_{\nu}}\right)  \\
&& -\frac{1}{2}\cos^4 \theta_{13}\sin^2 2\theta_{12}\left(1-\cos\frac{L\Delta m^2_{sol}}{2E_{\nu}}\right) \nonumber \\
+\frac{1}{2}&\sin^2 2\theta_{13}\sin^2 \theta_{12}&\left(\cos\frac{L}{2E_{\nu}}\left(\Delta m^2_{atm}-\Delta m^2_{sol}\right)
-\cos\frac{L\Delta m^2_{atm}}{2E_{\nu}}\right), \nonumber
\end{eqnarray}
(the probability for the inverted hierarchy $P_{IH}$ is obtained simply by replacing $\sin^2\theta_{12}$ with $\cos^2\theta_{12}$ in the last term above).
JUNO will be placed about 700 m below underground, corresponding to 
about ~1900 m.w.e., in a pit dug in the ground afresh. 
To pursue its main physics goals JUNO will need to attain 
an unprecedented resolution 
on the energy of the $\nubare$ produced in the reactors. 
In order to meet such performance extensive studies have been conducted
over the past few years concerning various aspects of engineering, 
detector design, including 
optical/light collection in the LS and PMT and response of the read-out electronics, 
software development and background suppression \cite{junoCDR}. 
From these developments also analyses of other aspects of neutrino physics will benefit, for example 
the measurement of the fluxes of solar neutrinos impinging on the JUNO detector and their energy distribution. 
This is the main subject of my talk given at the 
5$^{th}$ international solar neutrino conference. In this talk I first described the status of the detector design optimization and 
construction; then I illustrated its physics potential in the observation
of solar neutrinos specifically. The other parts of the rich physics programme of JUNO were not discussed in this topical conference. 

\section{What drives the detector design....}
The experimental signature of the reactor $\nubare$ in the JUNO detector
 is given by the inverse beta decay (IBD) process 
$\ensuremath{\overline{\nu}_e}\xspace + p \rightarrow e^+ + n$,
where $p$ and $n$ indicate a proton from the LS and a neutron, respectively.
The resulting signal is given by a prompt ``flash'' of light from the positron energy loss and 
annihilation; plus delayed light at a fixed 2.2 MeV energy from the neutron capture on proton.
The main goal of JUNO is to determine the MH with a significance of at least 3$\sigma$
within the first 6 years of data taking. The correct MH will be extracted by means 
of a $\chi^2$ fit to the kinetic energy spectrum
of the prompt $e^+$ ($T_{e^+}$), which is directly related to the 
$\nubare$ energy ($E_{\nubare} = T_{e^+} + 2\times m_e + 0.8$ MeV, where $m_e$ is the 
$e^+$ mass). 
From Eqn. \ref{eq:1} one sees that a difference arises in the fine structure of
the $\nubare$ energy spectrum, for the two mass hierarchy hypotheses, as was pointed out in Ref.\cite{petcov}. 
\begin{figure}[t]
     \centering
     \begin{center}
       \mbox{
         \epsfig{file=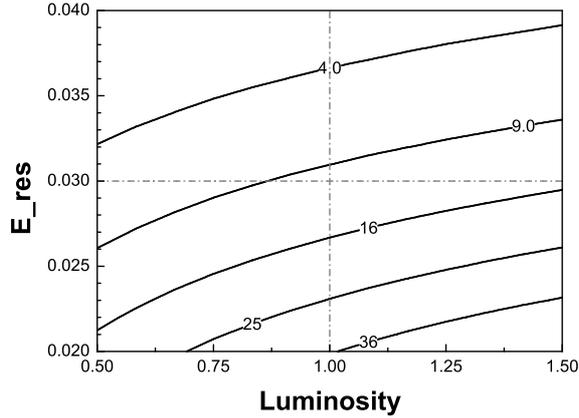,height=7 cm}
       }
     \end{center}
     \caption{Curves of $\Delta\chi_{min}^2 (NH-IH)$ as a function of the relative energy resolution $\delta E/E$ (y axis) and the ``luminosity'' of the sample collected with respect to the 6-year benchmark \cite{An:2015jdp}. Configurations of $\delta E/E$ and luminosity corresponding to the same value of $\Delta\chi^2$  are shown as solid lines, together with the corresponding value of $\Delta\chi^2$ testing two MH hypotheses.}
     \label{fig:1}
\end{figure}
The correct MH is determined by constructing the $\Delta\chi_{min}^2 (NH-IH)$ from 
the two fits to the experimental reactor data; this can be translated into a statistical 
significance of the discrimination.
It should be noticed that, with such strategy, a residual ambiguity lingers, associated to the correct 
value of the atmospheric mass difference $\Delta m^2_{atm}$; this reduces the 
final sensitivity of the fitting procedure. 
JUNO estimates that, in order to reach the desired significance, the most important figure of
performance is the overall resolution on the event-by-event measurement of $E_{\nubare}$.
In Fig.~\ref{fig:1} this relationship appears clearly. 
It is therefore crucial to design a detector which minimizes
the statistical uncertainty from stochastic fluctuations in the scintillation light collection, 
yet keeping linearity and uniformity of the energy response under control.
A 3$\%$ overall relative energy resolution will yield, for 6 years of data at 36 GW of 
reactor thermal power, a median significance of 3.4 (3.5) $\sigma$ for NH (IH) \cite{An:2015jdp}. 

\section{...and the resulting detector design}
\subsection{Central Detector}
To attain this level of precision, the JUNO collaboration has developed a detector made of 
3 basic parts, plus the electronics. The central detector (CD) is a 20 kton LS target mass, 
conceived to maximize the photon statistics and minimize the attenuation 
of the IBD prompt signal. This will be the largest volume of LS to date, composed of a mixture 
of $>98\%$ LAB (solvent, ~1200 photo-electrons/MeV), PPO (solute) and a less-than-per-cent 
part of bis-MSB (wavelength shifter). The photons are collected by PMT of two different kinds: 
about 18000 20 inch PMT, most of which of the micro-channel plate 
 (MCP-PMT) type, will guarantee an extended photo-coverage ($75\%$, as per JUNO requirement)
and a high overall detection efficiency (expected by the JUNO specifications to be $27\%$ at $\lambda$=420 nm); 
their transit time spread (TTS) being 12 ns. 25000 
“conventional” 3 inch PMT will allow to follow a multi-calorimetric approach, whereby the 
non-stochastic behaviour in the energy resolution will 
be monitored during the calibration runs with known radio-active sources at different 
energies; the 3 inch PMT will extend the dynamical range in the waveform of large numbers 
of photo-electrons hitting a localized region and improve time and vertex resolution 
for muon reconstruction (against cosmic muons). 
\subsection{Veto}
Additionally, a veto system will be in place to screen off incoming muons and photons by 
means of a surrounding water buffer (Cherenkov) and top plastic scintillators. 
The low energy secondary light from radioactive processes in the experimental cavern and from the detector material itself; plus particles from incoming cosmic muons represent the two main classes of backgrounds for the JUNO physics programme. They are particularly relevant for solar neutrino studies. 
\subsubsection{Water pool}
Even after a thorough purification campaign of the LS to the desired level for the given physics targets, radioactivity will come from not perfectly clean glasses and materials in the acrylic sphere, optical instrumentation and read-out systems. $\gamma$ from $^{40}$K, $^{214}$Bi and $^{208}$Tl could enter the active volume and need to be screened off effectively. Also, cosmic muons can interact with $^{12}$C in LS and produce lighter isotopes (esp. $^9$Li and $^8$He). Since the JUNO experimental cavern is placed at a relatively small depth of 700 m under ground, the muon rate in the CD is expected to be about 3 to 4 Hz, with an average muon energy of 215 GeV. In the surrounding water buffer pool the rate is expected to be approximately 1 to 2 Hz. A pool of 35 kton of ultra-pure water (water pool, WP) will be placed in a cylinder surrounding the CD sphere. Immersed in it, about 2000 20-inch PMT will collect the Cherenkov light produced by the traversing secondary particles like cosmic muons to identify and stop accompanying neutrons from the $^{12}$C scatter (active veto). At the same time the water will passively screen off the $\gamma$ particles from the rocks. The WP veto should be effective at more than 95$\%$ against these backgrounds. 

\subsubsection{Top tracker}
Finally, a ``top tracker'' (TT) using the plastic scintillator layers of the former OPERA experiment (at Laboratori Nazionali del Gran Sasso, Italy) will provide a geometrical coverage of about 50$\%$ to veto and/or collect a calibration sample of cosmic muons passing through the JUNO CD. The TT will allow to study the performance of dedicated tracking algorithms.  

\subsection{Front-end}
The project of the front-end 
electronics is also a challenge, because of the many read-out channels, the dark noise rate of the 
20 inch PMT and the needed resiliance and efficient heat dissipation of an under-water system. A large effort from the Collaboration is being devoted to this task.

\section{Current status of the detector project}
\begin{figure}[t]
     \centering
     \begin{center}
         \epsfig{file=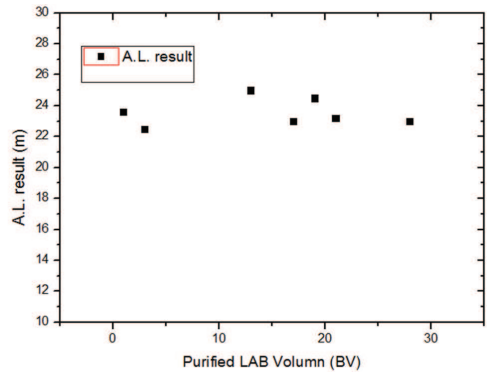,height=7cm}
     \end{center}
     \caption{Attenuation length vs level of LAB purification from the Daya Bay purification pilot plant.} 
     \label{fig:2}
\end{figure}

After a careful design phase, the construction of the various elements is underway. 
About 15000 MCP-PMT were ordered from the NNVT (North Night Vision of Technology CO., LTD, China) \cite{bib:nnvt}
manufacturer and are being tested at 
a dedicated centre in the region around the JUNO site. 
Additional 5000 “conventional dynode” 20 inch 
PMT were commissioned to Hamamatsu \cite{bib:hama}, of the type R12860, in order to complement the 
leading PMT lay-out and provide a faster TTS (3 ns). All the large-area PMT will be 
equipped with protective masks to prevent the generation of shock waves if one PMT 
explodes under water pressure: their design has been finalized after extensive 
pressure tests. The bidding of the 3 inch PMT was completed last spring and
 the elements are currently buing built by from HZC-Photonics. They are custom-made based on the KM3NeT
design, with improved TTS for better muon tracking. It is desirable that the LS be purified 
to a good degree from radioactive isotopes, to reduce the intrinsic background of
the detector. The required concentrations are $10^{-15}$ g/g for $^{238}$U and $^{232}$Th and 
$10^{-17}$ g/g for $^{40}$K. The attenuation length (AL) that JUNO requires is 
greater than 20 m at $\lambda=$ 430 nm (for 3 g/l of PPO in LAB). 
A strategy has been developed by JUNO aimed at obtaining an optimal 
admixture of solvent and
solutes in optical and radio-active terms. The purification will go through four parallel 
and complementary processes: an Al$_2$O$_3$ (alumina) column, a distillation plant, 
water extraction and gas stripping. 
A pilot plant has been established in one of the LS halls of the Daya Bay experiment
in China to monitor the AL and level of purification, which uses the alumina method. 
Results are displayed in Fig.~\ref{fig:2} and show a good stability and that the desired 
level of AL has been matched.

\section{Solar neutrinos in JUNO: the main strategy}
The main aims of our solar neutrino programme are to provide new measurements of $^7$Be and $^8$B. 

JUNO will select solar neutrino interactions as ``singles'' from the elastic electron scattering process in the LS: 
\begin{center}
\begin{equation}
\nu_{e} + e^- \rightarrow \nu_{e} +    e^-
\end{equation}
\end{center}
a weak current process with no energy threshold.
Using the liquid scintillator technique offers the advantage to be able to fill a large sensitive mass and guarantees to JUNO a big exposure. This is ideal to enhance the solar neutrino statistics, compared to its predecessors. The additional main benefit of JUNO for solar neutrino physics comes from its unprecedented calorimetric resolution. Importantly, this will allow to isolate the $^7$Be ``shoulder'' in the scattered electron energy spectrum to isolate it from radioactive backgrounds. 

On the other hand, a large ``monolithic'' detector {\it a la} JUNO offers no directionality and therefore no event-by-event background subtraction: just like in Borexino, the rejection of $\alpha$, $\beta$ and $\gamma$ will be on statistical grounds only. 

Finally, I have already commented on the depth of JUNO and the cosmic muon rate higher than for other, similar solar neutrino experiments: this might reduce our capabilities to $^8$Be (and perhaps also to hep neutrinos beyond 1 MeV energy).

\section{Main backgrounds}
\subsection{Reducing the backgrounds}
As a benchmark, I report in Tab.~\ref{tab:mh:sigbkg} the antineutrino selection efficiency for IBD signals. After the fiducial volume it is 91.8\%. An energy cut, time cut and vertex cut, described in Ref.\cite{An:2015jdp}, are applied to suppress the main classes of backgrounds: geo-neutrinos, cosmogenic and accidental associations of radioactive ``singles'' into a ``double'' signal like IBD.  
The cuts have efficiencies of 97.8\%, 99.1\%, and 98.7\%, respectively, using Geant4-based Monte Carlo (MC) studies. A muon veto cut will also be applied. Its efficiency is estimated to be 83\%, assuming 99\% muons have a good reconstructed track. 
After all cuts JUNO will observe 60 IBD events per day, with a background contribution to the total collected sample of about 6\%. Tab.~\ref{tab:mh:sigbkg} shows that the relative importance of the three classes will be similar after the muon veto. 

\begin{table}[!htbp]
  \tbl{The efficiencies of antineutrino selection cuts,
    signal and backgrounds rates. From Ref.\cite{An:2015jdp}\label{tab:mh:sigbkg}}{
    \begin{tabular}{|c|c|c|c|c|c|c|c|}\hline\hline
      Selection & IBD efficiency & IBD & Geo-$\nu$s & Accidental & $^9$Li/$^8$He & Fast $n$ & $(\alpha, n)$ \\ \hline
      - & - & 83 & 1.5 & $\sim5.7\times10^4$ & 84 & - & - \\ \hline
      Fiducial volume & 91.8\% & 76 & 1.4 &  & 77 & 0.1 & 0.05 \\ \cline{1-4}\cline{6-6}
      Energy cut & 97.8\% & & & 410 &  &  &  \\ \cline{1-2}
      Time cut & 99.1\% & 73 & 1.3 &  & 71 &  &  \\ \cline{1-2}\cline{5-5}
      Vertex cut & 98.7\% & & & 1.1 &  &  &  \\ \cline{1-6}
      Muon veto & 83\% & 60 & 1.1 & 0.9  & 1.6 &  &  \\ \hline
      Combined & 73\% & 60  & \multicolumn{5}{c|}{3.8} \\ \hline
      \hline
  \end{tabular}}
\end{table}

\subsection{Radioactive background budget}
Preliminary upper limits, obtained from laboratory measurements of the concentration of radioactive nuclides in the JUNO detector material,  indicate that one can expect about 60 Hz of ``single'' signals in the whole CD volume, while a fiducial volume of $r<17$ m (with $r$ the distance from the centre of the LS spherical tank) reduces that to about 8 Hz. Assuming these preliminary results, the largest sources of accidental background will be the LS, the acrylic containing it, the PMT glass and the radon in the WP. Even after the fiducial volume cut, the PMT glass would still be the dominant source of radioactive background (equally shared between the $^{238}$U and $^{232}$Th chains).

\subsection{Radio-purity scenarios}
Because radio-activity plays such an important role in the energy range of solar neutrinos, one needs to make assumptions on its intensity in order to extract the JUNO potential for solar neutrinos. Two scenarios of radio-purity in JUNO are assumed in the following: 
\begin{itemize}
\item ``baseline'': this is the minimum requirement believed to be necessary to do any solar neutrino analysis. The levels set as tolerable are summarized in Tab.~\ref{tab:solar:singles}. This level of radio-purity would give a signal-to-background yield ratio $S/B \approx 1/3$, about the same as attained in the KamLAND highest purity for the ``solar'' phase. It should be noticed that this means a factor 10 better than the purity level required for IBD (oscillation analysis), as in this case one relies on ``doubles'' to identify the signal.\\

\item ``ideal'': with an $S/B \approx 2/1$ this would be a level similar to what Borexino had in phase-I. 
\end{itemize}

\begin{table}[h!]
  \tbl{The requirements of singles background rates for doing low energy solar neutrino measurements  and the estimated solar neutrino signal rates at JUNO. Taken from Ref.\cite{An:2015jdp}: in that reference the last row had wrong (underestimated) values from the simulation, which have now been corrected. \label{tab:solar:singles}}{  
    \begin{tabular}{| r | c | c |}
      \hline \hline
      \multicolumn{3}{|c|}{Internal radiopurity requirements} \\
      \hline
      & baseline & ideal \\
      \hline
      $^{210}$Pb & $5\times10^{-24}$ [g/g] & $1\times10^{-24}$ [g/g] \\
      $^{85}$Kr  & $500$ [counts/day/kton] & $100$ [counts/day/kton] \\
      $^{238}$U & $1\times10^{-16}$ [g/g] & $1\times10^{-17}$ [g/g] \\
      $^{232}$Th & $1\times10^{-16}$ [g/g] & $1\times10^{-17}$ [g/g] \\
      $^{40}$K & $1\times10^{-17}$ [g/g] & $1\times10^{-18}$ [g/g] \\
      $^{14}$C & $1\times10^{-17}$ [g/g] & $1\times10^{-18}$ [g/g] \\
      \hline
      \multicolumn{3}{|c|}{Cosmogenic background rates [counts/day/kton]} \\
      \hline
      $^{11}$C & \multicolumn{2}{|c|}{$1860$}  \\
      $^{10}$C & \multicolumn{2}{|c|}{$35$}  \\
      \hline
      \multicolumn{3}{|c|}{Solar neutrino signal rates [counts/day/kton]} \\
      \hline
      pp $\nu$ &\multicolumn{2}{|c|}{$1378$ } \\
      $^{7}$Be $\nu$ &\multicolumn{2}{|c|}{$517$} \\
      pep $\nu$ &\multicolumn{2}{|c|}{$28$} \\
      $^{8}$B  $\nu$ &\multicolumn{2}{|c|}{$4.5$} \\
      $^{13}$N/$^{15}$O/$^{17}$F  $\nu$ &\multicolumn{2}{|c|}{$25 / 28 / 0.7$} \\
      \hline  \hline
    \end{tabular}}
\end{table}

The results of Tab.~\ref{tab:solar:singles} are obtained from simulations, using the BP05(OP) flux prediction \cite{bib:bahcall} for solar neutrinos, the current best values for the elastic scattering cross-sections, no energy threshold cuts and no fiducial volume cut. $^{10}$C and $^{11}$C are scaled from the KamLAND spallation measurements \cite{bib:kamland} considering the depth (scale factor = 0.9).

More realistic numbers will only be obtainable after the campaign of concentration and activity measurements of the various materials will be completed; and the simulation tuned accordingly for positioning of the material in the JUNO detector geometry. 

As a side remark, {\it in-situ} determinations of these numbers with first JUNO data will be an important additional constraint to get the final picture. 

\section{Towards $^7$Be}
\begin{figure}[!htb]
  \centering
  \subfigure[]
  {
      \label{fig:solar:simul2:a}
      \includegraphics[width=0.8\textwidth]{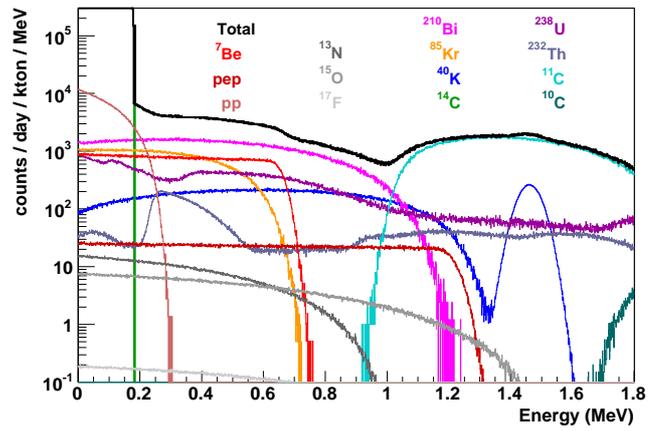}
  }
  \hfill
  \subfigure[]
  {
      \label{fig:solar:simul2:b}
      \includegraphics[width=0.8\textwidth]{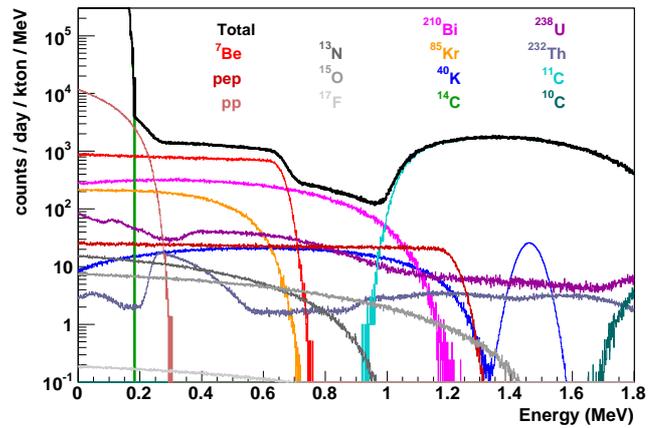}
  }
  \caption{The expected singles spectra at JUNO with (a) the ``baseline'' and (b) the ``ideal'' radiopurity assumptions listed in Table ~\ref{tab:solar:singles}. Taken from Ref.\cite{An:2015jdp}}
  \label{fig:solar:simul2}
\end{figure}

How rosy the prospects for $^7$Be are depends largely on the amount of background at low energies (mainly radioactive processes). The spectrum of the electrons from the elastic scattering is shown in Fig.~\ref{fig:solar:simul2} for the two radio-purity scenarios defined above. 
It can be seen that the situation, in the case of the ``ideal'' scenario, looks promising, with the features of the $^7$Be electron spectrum evident in the overall spectrum: they can be used in a fit-based extraction of the integrated flux.  
There are several caveats about how the distributions of Fig.~\ref{fig:solar:simul2} were obtained. First of all, here only the internal radioactivity of the LS is considered. That is to say, the incoming particles from radioactive decays in the surrounding materials and rocks are neglected: the spectra will be realistic only where, for example, a fiducial volume cut will be able to suppress most of this external background. Also, $^{238}$U and $^{232}$Th decay products are supposed to be created at saecular equilibrium: while this is a reasonable assumption, it remains to be seen whether any additional external ``polluting'' sources might break it, as it has happened for example with $^{210}$Pb in the Borexino vessel, whose daughters (namely $^{210}$Bi and, in turn, its daughter $^{210}$Po) diffuse into the LS by convective motions (for a description of the issue see, e.g., the talk at Ref.~\cite{bib:guffanti} at this conference).

In Fig.~\ref{fig:solar:simul2} no $\alpha$ peak from $^{210}$Po $\rightarrow$ $^{206}$Pb $+\alpha$ (taking the isotopes from the $^{238}$U chain as an example) are shown. This is a deliberate choice, which stems from the assumption that $\alpha$ particles can be successfully identified using Pulse-Shape Discrimination (PSD) techniques, thus suppressing the $^{210}$Po background \cite{bib:borexp1}. Only $\beta$ and $\gamma$ radiation is shown. 
In any case, taking into account quenching of heavier particles in the LS and the relatively high energy threshold of the detector (about 500 keV), it remains to be seen whether $\alpha$ events will be detected.

The normalization of $^{85}$Kr in the figure is set to the tolerable levels of Tab.~\ref{tab:solar:singles}. These levels assume no air leaks into the detector, which would carry $^{85}$Kr (and $^{39}$Ar); and use of ultra-pure water steam in the stripping process to remove gaseous (radio-)impurities.

Finally, Fig.~\ref{fig:solar:simul2} corresponds to the detector simulation level. The signal is shown after the simulation of light in the LS and optical collection at the PMT. The component of the uncertainty included there is only coming from photostatistics: no dark noise of the PMT, no distortions of the PMT signal nor overlapping of the $^{14}$C at the single PMT level are considered. 

\subsection{Hindrances: dark noise and $^{14}$\text{C} pile-up} 
Particularly two effects are relevant for solar neutrino physics: the addition of photo-electrons at the PMT photo-catode from ``spurious'' sources like dark thermal noise (DN) and the overlap of $\beta$ decays of the $^{14}$C, unavoidably present along the LS hydrocarbon, to signals. 
\subsubsection{Dark noise}
\begin{figure}[t]
  \begin{flushleft}
    \mbox{
      \epsfig{file=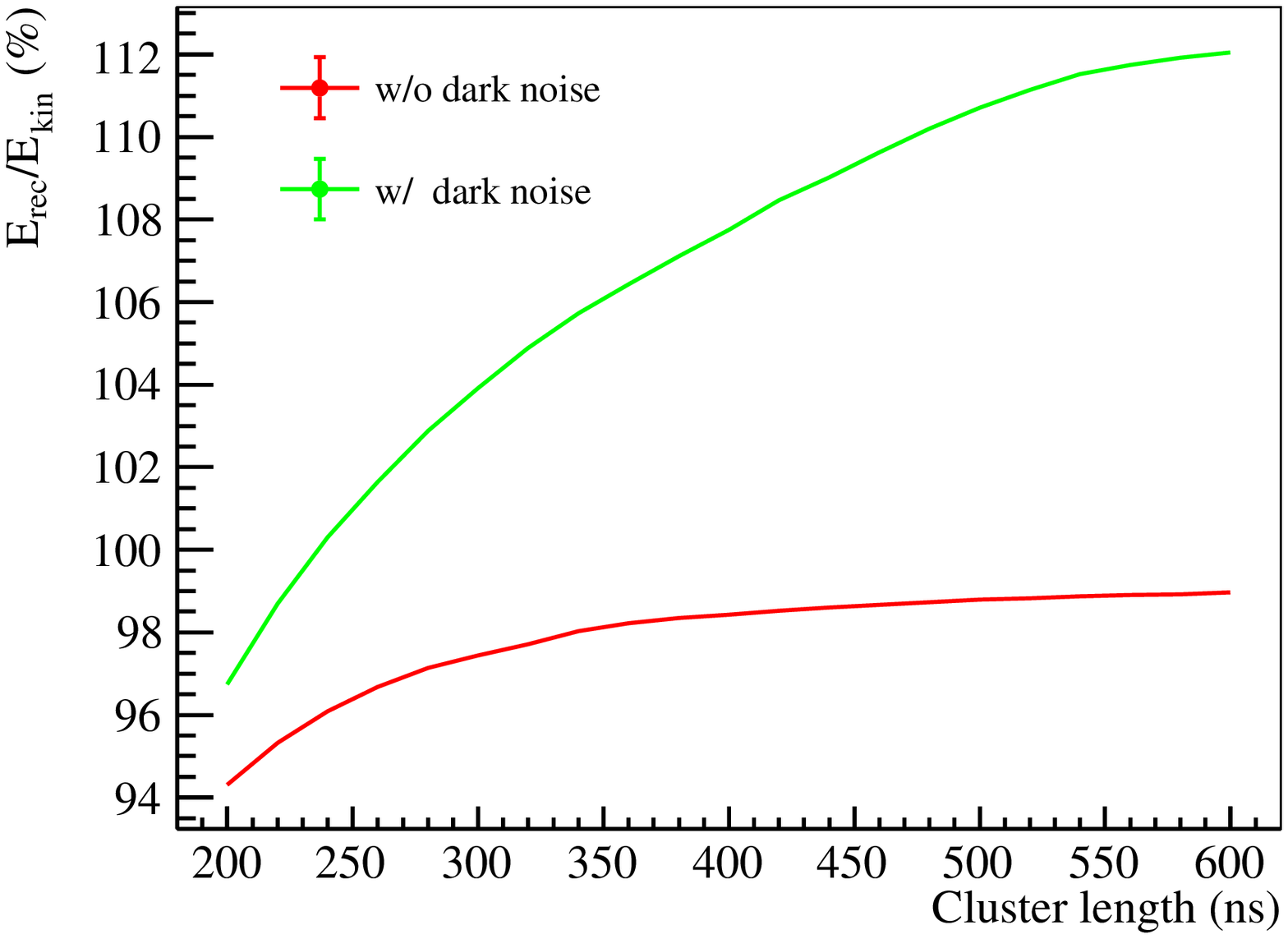,height=4.5 cm}
      \epsfig{file=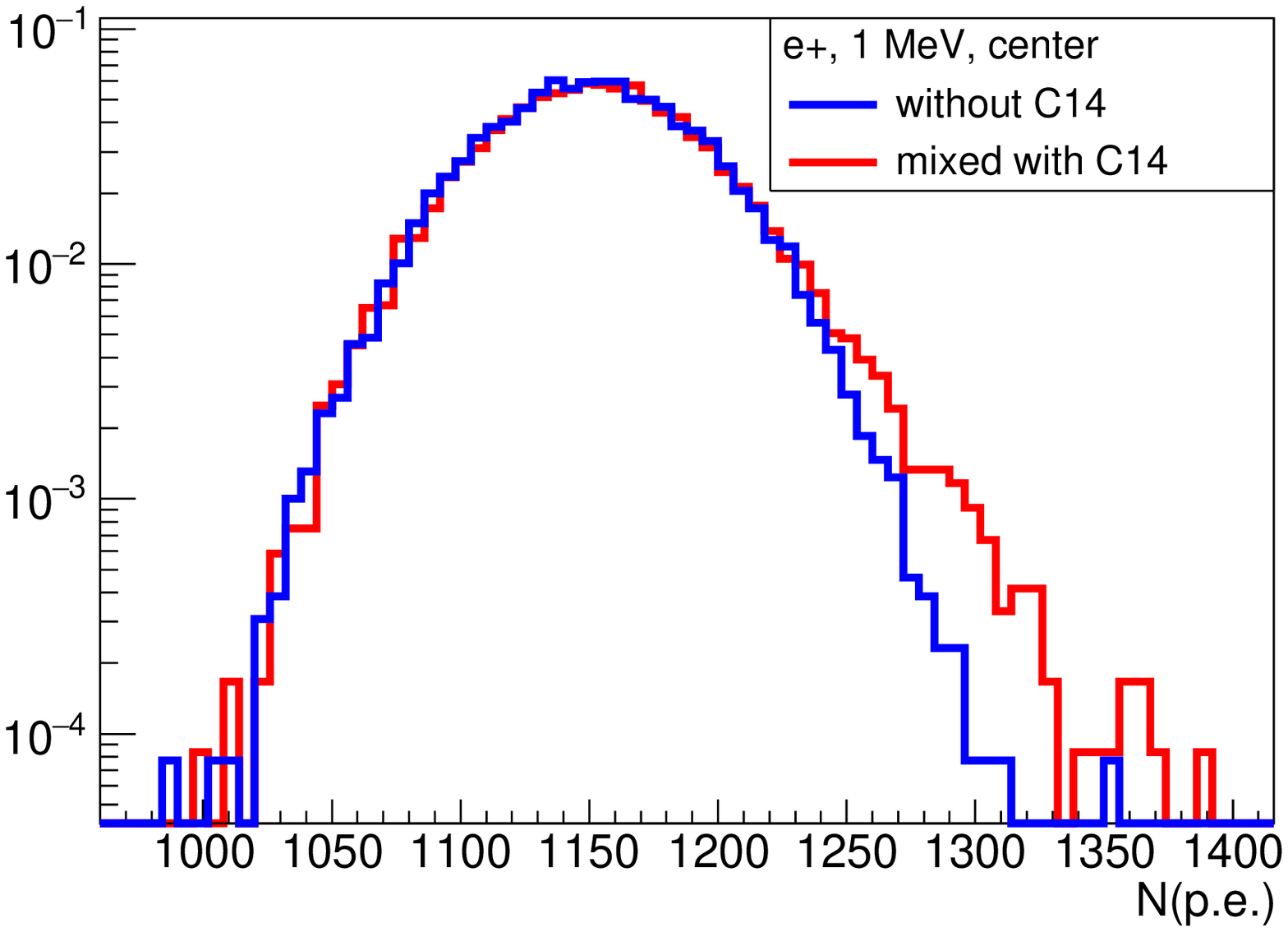,height=4.5 cm}
    }
  \end{flushleft}
  \caption{Left: Expected reconstructed energy devided by the true energy for 1 MeV electrons in the JUNO LS as a function of the cluster length: in red the case where no dark noise hits overlap with the electron energy deposition; in green when dark noise hits contribute to the visible energy ({\it courtesy X. Ding, shown at Neutrino 2018 conference} \cite{bib:ding}). Right: number of photo-electrons generated on the JUNO optical system by 1 MeV single positron events depositing energy in the LS (blue); and the same with the contribution of $^{14}$C. ({\it courtesy P.Kampmann, shown at the DPG2018} \cite{bib:kampmann})}
  \label{fig:be7-biases}
\end{figure}

The JUNO PMT will have a large quantum efficiency, but that also carries the downside of sizeable dark currents. The average dark rate of the PMT will be of several tens of kHz. While it will be possible to select signal events thanks to time coincidences of several PMT signals, the danger is that the energy (in terms of number of PMT or number of photo-electrons on those PMT, whichever estimator is chosen) will be misestimated because of the DN contribution. This is particularly relevant for low energy electrons as those from $^7$Be. Not only this will impact the energy linearity, but it will also spoil the effort to maintain the desired 3$\%$ relative resolution around 1 MeV. Clustering algorithms were already successfully applied in Borexino \cite{bib:borexp1}, where only hits ``clustered'' around a given time (for example the trigger time) are considered and the rest is rejected. This is also being optimized for JUNO. A preliminary idea of how much this algorithm can help is shown in the left panel of Fig.~\ref{fig:be7-biases}, where the residual energy bias from the DN contribution is shown as a function of the time window considered to define a cluster. 

\subsubsection{$^{14}$\text{C} pile-up}
Assuming a ratio of $^{14}$C to the stable $^{12}$C in the LS to $10^{-17}$ and a half-life of $^{14}$C $T_{1/2}=0.25\times10^{12}$ s, the total activity of $^{14}$C in the LS is 40 kBq. With such an occurrence, one estimates as 5$\%$ the probability that an uncorrelated $^{14}$C decay produces hits overlapping to a signal of arbitrary rate within a 1250 ns data acquisition time window (as currently expected for JUNO).
The plot on the right of Fig.~\ref{fig:be7-biases} shows what the result would look like for 1 MeV positrons. As can be seen, this translates into a bias on the light yield towards a high tail in the number of photo-electrons generated on the JUNO 20'' PMT. This will be even more relevant for solar neutrino signals. The same clustering algorithm is expected to largely suppress $^{14}$C hits, as these have a uniform time pattern. 

\section{Prospects for $^8$B}
In the energy region where the elastic-scattered electrons from $^8$B are dominant (from a few MeV on) and the fit can be most sensitive to this component, the effect of cosmogenic backgrounds will be most important. These will come from the spallation of cosmic muons on $^{12}$C nuclei contained in the LS molecules. This ultimately gives rise to radioactive decays of various isotopes. However, we assume that only the 3 longest lived ones will represent an issue: $^{10}$C, $^{11}$C and $^{11}$Be. Their characteristics are recalled in Tab.~\ref{tab:solar:cosmogenic_isotopes}

\begin{table}[!h]
  \tbl{List of the cosmogenic radioisotopes whose lifetimes are above 2\,s, which are the main backgrounds for $^8$B solar neutrino detection. The table is taken from Ref.\cite{An:2015jdp}\label{tab:solar:cosmogenic_isotopes}}{
    \begin{tabular}{|c|c|c|c|c|c|}
      \hline
      Isotope& Decay Type & Q-Value  & Life time & Yield &Rate \\
      &   &  [MeV] &  & $10^{-7}$ $(\mu \, {\rm g}/{\rm cm}^2)^{-1}$ & [cpd/\,kton]\\
      \hline
      $^{11}$C& $\beta^+$ & 2.0 & 29.4\,min & 866 & 1860\\
      $^{10}$C& $\beta^+$ & 3.7 & 27.8\,s & 16.5 & 35\\
      $^{11}$Be& $\beta^-$ & 11.5 & 19.9\,s & 1.1& 2\\
      \hline
    \end{tabular}}
\end{table}

\begin{figure}[!tb]
\vspace{0 cm}
\begin{center}
\includegraphics[width=0.7\textwidth]{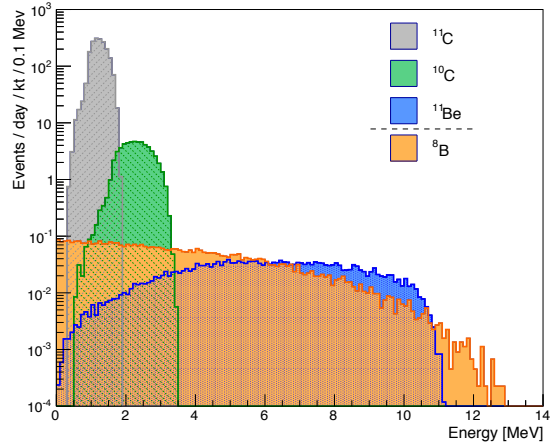}
\vspace{0 cm}
\end{center}
\caption{The simulated background spectra for the cosmogenics isotopes $^{11}$C, $^{10}$C, and $^{11}$Be at JUNO. Furthermore, the expected $^8$B ($\nu_e$) spectrum is shown for comparison.}
\label{fig:solar:boro}
\end{figure}

Signals from short-lived isotopes ($\tau \geq 1$s) can be effectively targeted and suppressed if a good muon tracking is in place. Preliminarly JUNO will be able to reject those events by vetoeing light from a cylindrical volume around the muon track in the LS. Preliminary selections are set to a cylinder of $r=$1 m for a time of 6.5\,s, which keeps the dead time of the experiment at a manageable level. 
However, vetoeing a region for the time longer lived isotopes take to decay (up to about 30 minutes for $^{11}$C) is not feasible in terms of dead time. The idea is to fit them together with the $^8$B electron spectrum. A preliminary estimation of the overlaid spectrum of signal and above backgrounds, from simulation, is shown in Fig.~\ref{fig:solar:boro}.

A second idea is being pursued to detect $^8$B neutrino interactions and measure their energy with a much reduced background (which gives ``singles''). It relies on identifying $^8$B neutrinos via charged current events instead of elastic scattering on electrons. In this case the process: 
\begin{center}
\begin{equation}
\nu_{e} + ^{13}\text{C} \rightarrow  e^- + ^{13}\text{N} 
\end{equation}
\end{center}
is exploited, where the Q-value is 2.2 MeV and therefore the ``prompt'' electron has a kinetic energy exactly 2.2 MeV lower than the neutrino energy (ignoring the nucleus recoil). 
The subsequent $\beta^+$ decay of the $^{13}$N nucleus provides an ``IBD'' like double signature. The $^8$B signal can be selected from the energy loss of the initial electron and from the positron of the delayed decay (whose mean lifetime is $\tau=$862.8 s). Work is being developed to identify such correlated events to reduce the cosmogenic background by about 2 orders of magnitude from 2 MeV on in the electron kinetic energy. 

Concerning radioactive processes acting as background for $^8$B, the main source in the relevant energy range are the $\beta^-$ decays of $^{208}$Tl coming from the decay chain of $^{232}$Th present in the LS and in the PMT glass. The $\tau$ is of about 3 minutes and the Q-value of 5 MeV. This might become an issue especially if saecular equilibrium is broken and the 8.9 MeV $\alpha$ peak of the $^{212}$Po were not sufficient to estimate the $^{212}$Bi and the related $^{208}$Tl. 

Our expectations are that the rate of $^{232}$Th should be kept to below 10$^{-17}$ g/g (``ideal'' scenario) to be able to appreciate $^8$B at elastic-scattered electron energies below 5 MeV. It remains to be seen whether a fiducial volume cut (of up to 5 m from the acrylic vessel radius) will need to be applied. 

\section{Conclusions}
The JUNO experiment is on course to start operations within the next few years. 
JUNO will have two main advantages in the study of solar neutrinos: its large sensitive mass and its excellent energy resolution. Its capability to collect large statistics of elastic solar neutrino interactions and its ability to reconstruct the scattered electron spectrum accurately will allow JUNO to contribute significantly to the measurement of the $^7$Be and $^8$B fluxes, in the footsteps of previous experiments of similar technology. In the text initial expectations on the yields and spectra recontructed by JUNO for solar neutrino signals are provided. 
JUNO's relatively shallow position and radio-purity of its material and surrounding environment are the main challenges for solar neutrino measurements.  
A comprehensive programme is being put in place to keep the radio-activity to manageable levels. Algorithmic developments to identify and reject cosmogenic and intrisic backgrounds (like dark noise and $^{14}$C) are on-going. 
First data will allow JUNO to measure the actual levels of backgrounds and validate the assumptions made (for example whether any additional radioactive backgrounds are present in the experimental system beyond what expected from saecular equilibrium).
In my talk at the 5$^{th}$ International Solar Neutrino Conference I tried to put together a candid discussion of these aspects, highlighting advantages and issues for solar neutrinos at JUNO. A significant effort is being devoted to design detection strategies for $^7$Be and $^8$B. The feasibility of extracting CNO and hep neutrino fluxes is also being evaluated through novel ideas and with as realistic as possible simulations of the JUNO detector. 

\section{Acknowledgments}
I would like to thank the Organizers for the invitation and for putting together a lively conference of human-sized proportions. The well-structured sessions with top speakers in the various topics, the ample room for discussions and the moving historical reconstructions made this a truly cherishable moment. The beauty of Dresden made the rest.

\bibliographystyle{ws-procs9x6} 
\bibliography{salamanna_giuseppe}

\end{document}